\documentclass{article}

\usepackage{graphicx}
\usepackage{amsfonts} 
\usepackage{amsmath} 
\usepackage{amssymb} 
\usepackage{verbatim} 
\usepackage{algorithm}  
\usepackage{algorithmic}   
\usepackage{booktabs} 
\usepackage{amsthm} 
\usepackage[T1]{fontenc}
\usepackage[utf8]{inputenc}
\usepackage{authblk}

\newtheorem{remark}{Remark}

\title{Computation Load Balancing Real-Time Model Predictive Control in Urban Traffic Networks}
\author[a]{Qiming\ Zou}
\author[b]{Ke\ Lu \thanks{Corresponding author: luke.airoot@gmail.com}}
\author[b]{Yu\ Li}
\affil[a]{Department of Computer Science and Technology, Harbin Institute of Technology, China}
\affil[b]{Department of Management Science and Engineering, Anhui University of Technology, China}
\date{} 

\begin{document}
\maketitle
\begin{abstract}
    \looseness=-1 Owing to the rapid growth number of vehicles, 
    urban traffic congestion has become more and more severe in the last decades.
    As an effective approach, Model Predictive Control (MPC) has been applied to urban traffic signal control system.
    However, the potentially high online computation burden may limit its further application for real scenarios.
    In this paper, a new approach based on online active set strategy is proposed to improve the real-time performance of MPC-based traffic controller by reducing the online computing time.
    This approach divides one control cycle into several sequential sampling intervals. 
    In each interval, online active set method is applied to solve quadratic programming (QP) of traffic signal control model, by searching the optimal solution starting at the optimal solution of previous interval in the feasible region.
    The most appealing property of this approach lies in that it can distribute the computational complexity into several sample intervals, instead of imposing heavy computation burden at each end of control cycle.
    The simulation experiments show that this breakthrough approach can obviously reduce the online computational complexity, and increase the applicability of the MPC in real-life traffic networks.

\end{abstract}
\section{Intruduction}

    As the rapid increasing demand of urban traffic, congestion become much more serious than that of the past decades \cite{Zhou2017}, \cite{Luo2016}.
    Therefore, traffic control systems are installed to improve the performance of the existing urban transportation infrastructure, and thus to alleviate traffic congestion.
    Traffic-responsive control strategies which utilize the observed transportation states as feedback information provide an effective control approach for improving the performance of the transportation services in cities.

    
	 Traffic-responsive control strategies, including optimization-based and especially Model Predictive Control ($MPC$) can be used to manage the road capacity.
	 MPC  is robust to disturbances for signal control system\cite{JAMSHIDNEJAD2016147}, \cite{Jamshidnejad2015}.
	Additionally, as an optimal control approach, 
	$MPC$  can handle the intersection correlation as several constraints on states and the inputs \cite{Kothare2002Robust}, \cite{Muske2010Model}.
 	These characteristics make $MPC$ fit well for adjusting traffic signal control strategy to alleviates congestion.
	However, traffic-responsive approach need to rapidly adjust according to the real time traffic flow changes\cite{Ye2017}. 
	Hence, the repeated online computation may limits the application of $MPC$  in the large scale traffic system\cite{Bayat2011}, \cite{Reese2016}.
	
	In order to execute $MPC$ in millisecond range, a lot of attempts, which can be roughly classified into four types, have been proposed.
	Firstly, deriving explicit control law offline and searching the look-up table online, $e.g.$, Explicit Model Predictive Control ($EMPC$)  \cite{Bemporad2002}, \cite{Oberdieck2017}, \cite{LU2017}. 
	Secondly,  dividing the network into small subnetworks and executing the algorithms distributedly  \cite{Ocampo-Martinez2012}, \cite{Koehler2015}, \cite{Lin2015}, \cite{Zhou2017}.
 	Thirdly, simplifying the predictive model \cite{Lin2012}.
	Finally, increasing the efficiency of online computation \cite{Gould2017}, \cite{Jamshidnejad2017}.    
	Belonging to the final one, this work focuses on speeding up the convergence rate of $MPC$ by reducing the active set changes to improves the online computation of signal control system.	
	
	In this work, signal control problem is attributed in the form of quadratic programming ($QP$) under the $MPC$ framework, 
	and then solved in the way of obtaining the optimal solution.
    Active-set methods is one of the feasible approach to solve $QP$ problem and have advantage in hot start search \cite{4738961}.
    A great deal of research has improved the technology of warm-start for active-set methods.
    Based on muti-parametric programming, 
    Zeilnger computes the approximation of optimal solution offline  to warm-start the online computing \cite{Zeilinger2011}.
    Wang  uses the previously computed plan to obtained a predictive solution by  suitably shifted the previous solutions  in time, 
    and takes the predictive solution as a new starting point for the current plan \cite{Wang2010}.
    However, if the dynamical model is inaccurate, 
    both the predictive and previous solutions can't performance well enough to accelerate the iteration process. 
    Aiming this problem, 
    an efficient method called Online Active Set Strategy (OASS) is proposed to overcome this difficulty by taking advantage of the $QP$ solution of previous iteration \cite{Ferreau2007}, \cite{Ferreau2008}, \cite{Ferreau2014}.
    Different from using predictive solution as start point, OASS calculates the optimal solution along a straight line in parametric space, which starts from the previous QP to the current one.   
       Although, these methods are beneficial for optimization problem to converge faster, they can not well trade off between time and optimality.

	In this paper, we proposed a framework to improve the real-time feasibility without lose of optimality.
	In the traffic control system, the strategy is recalculated at intervals of one or several cycles which is called responsive time.
	It is a great waste that the control system have to stand idle until the end of responsive time when the traffic state is obtained.
	In this work, we will start calculating before the end of responsive time and gradually approaching to the optimal solution.
	This work stands out among other  approaches for three reasons.
	First, since this framework will output the suboptimal solution along the changes of state during responsive time, the computation complexity will be distributed to the whole time duration rather than one time point.
	Second, between two intervals, the state will not change significantly which will arise an appealing speedup using OASS.
	Last but not least, if too many active set changes are necessary to get from the solution of the old state to that of the current state,  OASS can just stop the solution of the current QP and start a new homotopy towards the solution of the next QP.
	This advantage gives our framework robustness to the external disturbance of the traffic dynamics.

	The rest of this paper is organized as follows.
	In Section 2, we describes the store-and-forward model, the formulation of MPC problem and the load unbalance problem of traditional traffic control system.
	Section 3 briefly summarized the main idea of  online active set strategy, and the main design procedure of the new framework.
	Afterwards, the experimental setup and the obtained results are described in Section 4;  
	Finally, the conclusion is drawn and our future work will be declared in Section 5.
\section{Problem Definition}

\subsection{Flow model}
  As to the research issues of transportation, a Urban Traffic Network(UTN) is usually deemed as a tuple consisting of a link set $S_{link}$ and a node set $S_{node}$.
    The links belonging to $S_{link}$ indicate roads, which are created by the intersections denoted by the nodes in $S_{node}$.

\begin{figure}[htbp]
        \centering
        \includegraphics[width=0.8\textwidth]{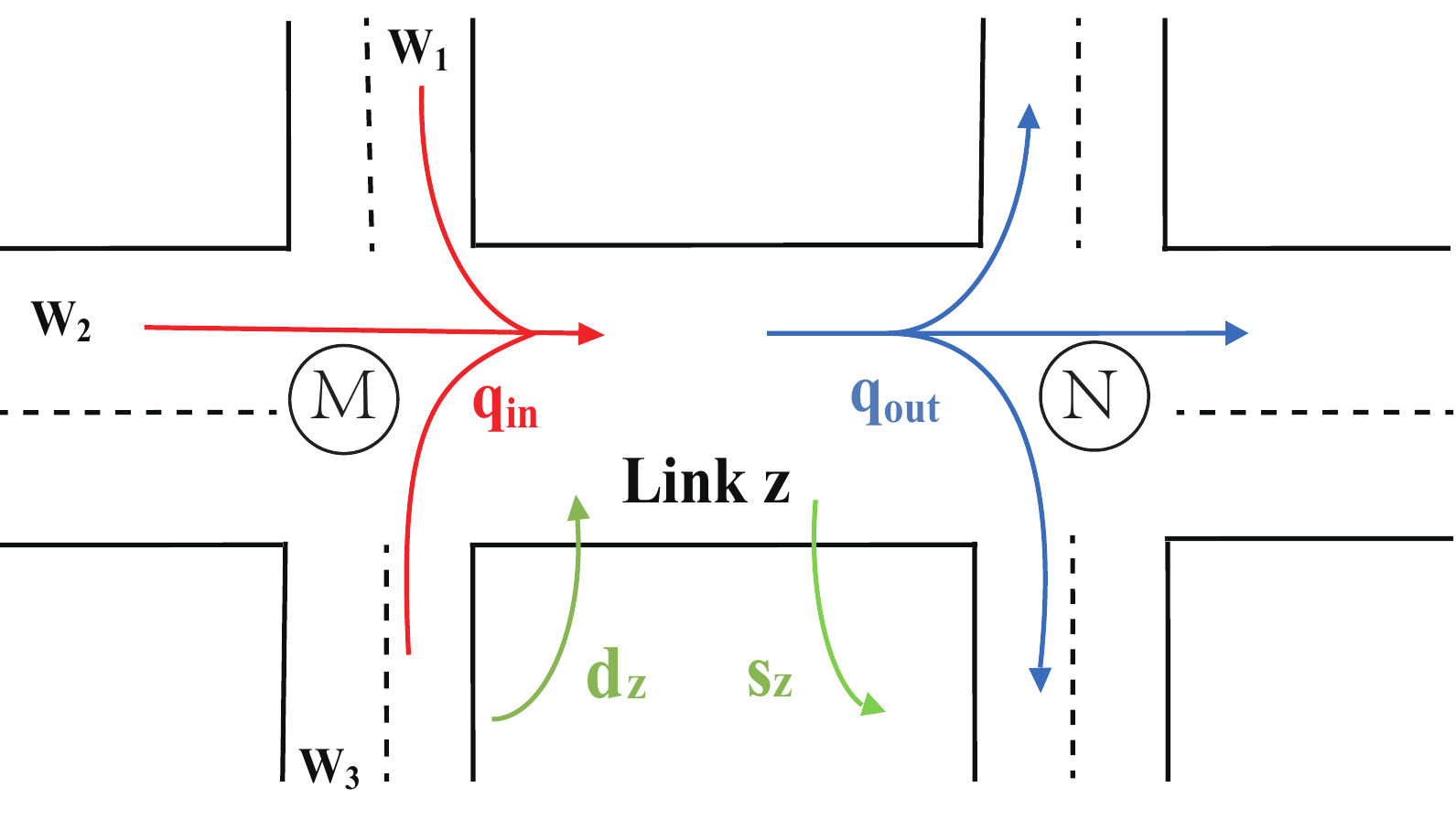}
        \caption{Traffic flow in the link z.}\label{fig:roadlink}
    \end{figure}

    In this paper, store-and-forward(SFM) model is selected to depict the variation of traffic states , e.g., the traffic flow, the turning rates and so on.
    This model paradigm is simple enough to be understood, and convenient for optimization of transportation system, e.g., the signal split problem discussed in this paper.
    The detail of this model are illustrated based on Fig.\ref{fig:roadlink}.
    
       According to Figure. \ref{fig:roadlink}, the $link~z$ represent the road between junctions of $M,N$.
    In this paper, the traffic flow dynamics of the link $z$ is formulated as equation(\ref{equation_dynamics})
    \begin{equation}\label{equation_dynamics}
      x_{z}(k+1)=x_{z}(k)+C[q_{in,z}(k)-q_{out,z}(k)+d_{z}(k)-s_{z}(k)]
    \end{equation}

    where $x_z(k)$ denotes number of vehicles in link $z$ at time step $k$;
    $q_{in,z}(k)$ and $q_{out,z}(k)$ represent the inflow rate  and the outflow rate of link $z$ respectively; $d_z$ is the demand flow and $s_z$ is the exit flow during the time interval $[kT, (k + 1)T]$, which are usually regarded as random  perturbation;
        $k$ denotes the discrete time step index and $T$ is the discrete time step.

    Before the further research on equation (\ref{equation_dynamics}), an critical variable ,$i.e.$,  the green time  vector $u$, need to be introduced.
    Firstly, every junction $j\in J$ has a signal control plan $u_j$. Furthermore, $u_j$ is based on a fixed number of phases that belong to the set $F_j$. Finally, the green time of phase $i$ at junction $j$ can be represented as $u_{j,i}$.

    For simplicity, the sampling time $C$ is equal to time step $T$ and the saturation flow rate $s_z$ is assumed to be known.
    An average value of $q_{out,z}(k)$ is obtained by
    \begin{equation}\label{equation_qout}
      q_{out,z}(k)=\frac{s_{z}G_{z}(k)}{T}
    \end{equation}
    $G_z(k)$ is the green time of link $z$, calculated as $G_{z}(k)=\sum_{i\in F_{N}}u_{N,i}(k)$.

    The inflow rate is given by
    \begin{equation}\label{equation_qin}
     q_{in,z}(k)=\sum_{w\in W_{z}}\tau_{w,z}q_{out,w}(k)
    \end{equation}
    where $W_z$ is the set of links which can lead to link $z$; the turning rates $\tau_{w,z}$ are representing the relative fraction of vehicles in link $w$ turning to link $z$.
    
    Replacing (\ref{equation_qin}) and (\ref{equation_qout}) in (\ref{equation_dynamics}) and generalize for all links in traffic network, the following matrix equation is derived.
    \begin{equation}\label{equation_dynamicsMatrix}
      x(k+1)=Ax(k)+Bu(k)+e(k)
    \end{equation}
    where
    \begin{equation}
   	x(k)=[x_1(k),...,x_n(k)]^T
   \end{equation}
    \begin{equation}
    u(k)=[u_{1,1}(k),...,u_{1,|F_1|}(k),...,u_{n,1}(k),\\...,u_{n,|F_n|}(k)]
    \end{equation}

\subsection{Optimization problem}
   It is necessary to point out that, our approach is not sensitive to the disturbance of model.
    Therefor, some disturbance factors that may affect the traffic flow,
    $e.g.$, the demand flow $d_z$ and the exit flow $s_z$ within the link $z$ rather intersections are ignored in the model for simplicity.
    In this paper, the following discrete-time linear time invariant system is considered.
   \begin{equation}\label{LTI_system}
    \left\{\begin{array}{ll}
    x_{t+k|t} = Ax_{t+k-1|t}+Bu_{t+k},\\
    y_{t+k|t} = Cx_{t+k|t},
    \end{array}
    \right.
    \end{equation}
    while fulfilling the constraints
    \begin{equation}\label{box_constraints_y}
    y_{min} \leq y_{t+k|t} \leq y_{max}
    \end{equation}
    \begin{equation}\label{box_constraints_u}
    u_{min} \leq u_{t+k} \leq u_{max}
    \end{equation}
    at all time instants $t \geq 0$.

    $x_{t+k|t}\in \mathbb{R}^n$ denotes the predicted state vector at time $t+k$,
    obtained by applying the input sequence $u_t,...,u_{t+k-1}$ to equation \ref{equation_dynamicsMatrix} starting from the state $x_{t|t}=x(t)$.
    $u_{t+k} \in \mathbb{R}^m$, and $y_{t+k|t} \in \mathbb{R}^p$ are the input, and output vector.

    \begin{equation}\label{equation_Tcons}
      \sum_{i\in F_{j}}u_{j,i}+L_{j}\leq T
    \end{equation}
    holds at each junction $j\in J$, where $L_j$, namely, lost time, is time span of all-red stages. Inequality (\ref{equation_Tcons}) is useful in cases of strong network congestion to allow for all red stages.
        
    Since the value of $x(t)$ may not be directly measurable but it is completely observed in this paper, we assume that the output vector $y(t)$ is equal to the state vector $x(t)$.
    Therefore, the inequation (\ref{box_constraints_y}) can be replaced by
    \begin{equation}\label{box_constraints_x}
    x_{min} \leq x_{t+k|t} \leq x_{max}
    \end{equation}

    Intending to minimize the risk of over saturation and spillback, minimization of proportion occupancy of links is attempted.
    To this end, the cost function has the form.
    \begin{equation}
    J(U,x(t))=x_{t+N_x|t}^TPx_{t+N_x|t}+\sum_{k=0}^{N_x-1}[x^T_{t+k|t}Qx^T_{t+k|t}+u^T_{t+k}Ru_{t+k}]
    \end{equation}
    Therefore, the online optimization problem of MPC with horizon N can be expressed as  
    \begin{equation}
     \label{optimization_problem_complex}
    \begin{split}
    & J^{*}(x(t))=\min_{U} J(U,x(t))\\
    & st.   x_{t+k+1|t}=Ax_{t+k|t}+Bu_{t+k}, 0 \leq k < N\\
    & x_{min}\leq x_{t+k|t}\leq x_{max}, 0 \leq k < N\\
    & u_{min}\leq u_{k+t}\leq u_{max}, 0 \leq k < N\\
    & \sum_{i\in F_{j}}u_{j,i}+L_{j}\leq C\\
    & x_{t|t}=x(t)\\
    \end{split}
    \end{equation}
where $Q$ and $R$ are the appropriately selected weighting matrix on the state and input.
    The diagonal elements of $Q$ is $1/x_z^{max}$, where $x_z^{max}$ is the greatest number of vehicles in the link $z$.
    We set this to normalize the number of vehicles because $x_z^{max}$ is different each link.
    This setting enable to put the weight to the links in which many vehicles exist in current time and distribute preferentially longer green time to the links.
    Matrix $R$ reflects the penalty imposed on control effort, it is set to be smaller than $Q$ as 0.01 $\times$ $I$ in order to consider the alleviation of traffic congestion to be important in this paper.
    
    The equaiton (\ref{optimization_problem_complex}) can be posed as a quadratic program in a standard form. By substituting
    \begin{equation}\label{equation_xN}
      x_{t+k|t}=A^{k}x_{t}+\sum_{j=0}^{k-1}A^{j}Bu_{t+k-1-j}
    \end{equation}
    equaiton (\ref{optimization_problem_complex}) can be rewritten as
    \begin{subequations}\label{optimization_problem}
    \begin{gather}
    J^{*}(x(t))=\frac{1}{2}x^T(t)Yx(t)+min(\frac{1}{2}{U}^THU+\nonumber\\U^Tg(x(t)))\\
    st. GU\leq b(x(t))
    \end{gather}
    \end{subequations}

    where the column vector $U \triangleq {[{u}^T_{0},...,{u}^T_{N_x-1}]}^T \in \mathbb{R}^s$, $s \triangleq mN_x$, is the optimization vector.
    $H=H^T \succ 0$, and $H,F,Y,G,W,E$ are easily obtained from $Q,R$,
    and equaiton (\ref{optimization_problem_complex})(as only the optimizer $U$ is needed, the term involving $Y$ is usually removed from \ref{optimization_problem}).

The gradient and constraint vector depend affinely on the initial value x(t): $g(x_0)=Fx(t)$ and $b(x(t))=W+Ex(t)$ for some matrix $E,F$.
    
MPC, as a traffic-responsive algorithm, needs to receive the measured states of the system at each end of control cycle.
The flow model will predictive the future states of the system based on the real current states, and the controller will minimize the cost function that is formulated based on the current and the predicted states. 
The output of the MPC controller is a optimal control sequence within the time span $[tT_c, (t+N)T_c)$, where $T_c$ denotes the length of the control cycle. 
Only the optimal strategy within $[tT_c, (t + 1)T_c)$  will be implemented to the real system. 
At time step $t + 1$, the states of the system will be measured again and will be used to calculate the optimal solution within $[(t+1)T_c, (t+N)T_c)$ in the same rolling horizon way. 

\subsection{Load unbalance problem}
\begin{figure}[htbp]
        \centering
        \includegraphics[width=0.7\textwidth]{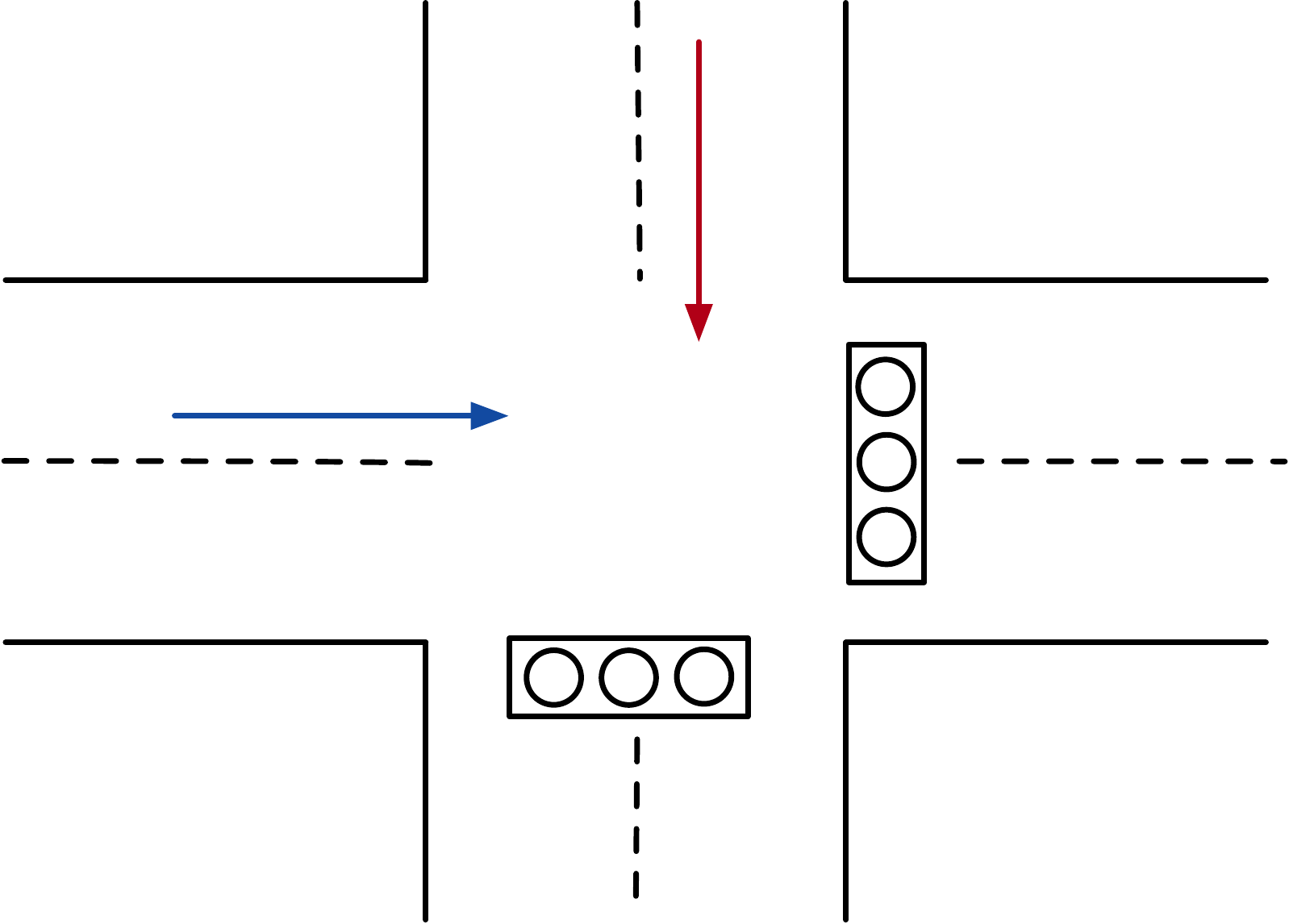}
        \caption{A two-dimension traffic signal control system. For simplicity, each signalized junction has only one phase, $i.e.$, link 1: north-to-south; link 2: west-to-east).}\label{fig:2Dexample}
\end{figure}  
The rolling horizon style introduced above makes MPC more robust towards external disturbances, while leads to a load unbalance problem. 
As the formulation of MPC, the state $x_t$ is measured at each end of control cycle, $i.e.$ the time point $tT_c$.
and the optimal solution is expected to be obtained at meanwhile. 
However, this kind of mechanism  imposes heavy computational loads upon the time point $tT_c$.

In order to illustrate this specific problem in traffic signal control scenario, a simple two-dimensional traffic system shown in Figure \ref{fig:2Dexample} is considered.

This system samples the dynamic with $T_s=T_c$ which is equal to 60s, and the state-space representation of the flow model is 
\begin{equation}
\label{eq:2d_exa}
\begin{split}
& x(k+1) = \begin{bmatrix}1 & 0\\ 0 & 1 \end{bmatrix}x(k)+\begin{bmatrix}-0.48 & 0\\0 & -0.48\end{bmatrix}u(k) +\begin{bmatrix}38\\38\end{bmatrix}\\
& y = \begin{bmatrix}1 & 0\\ 0 & 1 \end{bmatrix}x \\
\end{split}
\end{equation}
where $x = [x_1,x_2]^T$ is the predictive queue length of each road, $u = [u_1,u_2]^T$ is the green time implemented to each road. 
The task is to minimize Total Time Spend (TTS)\cite{Lin2012} while fulfilling the input constraints
\begin{equation}
5 \le u \le 55
\end{equation}
\begin{equation}
u_1+u_2 \le 60
\end{equation}
and the state constraints
\begin{equation}
0 \le x_1 \le 140
\end{equation}
\begin{equation}
0 \le x_2 \le 110
\end{equation}
To this aim, we design an MPC controller based on the optimization problem (\ref{optimization_problem_complex}).
Using a MPC controller, active set changes are 62, 64, 61, 65 times, and they are taken place at each end of cycle, $i.e.$, the time point 60, 120, 180, 240, 300, which shows the problem of load unbalance.

In the following section, we will  firstly introduce the main idea of online active set strategy, and describe how we can utilize the parametric property of this algorithm to distribute the calculation burden  into the entire control cycle.

\section{Methodolegy}

\subsection{Online active set strategy}
    In this subsection, the main idea of OASS is presented which was developed by Ferreau \cite{Ferreau2008}.

\begin{figure}[htbp]
\centering
\includegraphics[width=0.83\textwidth]{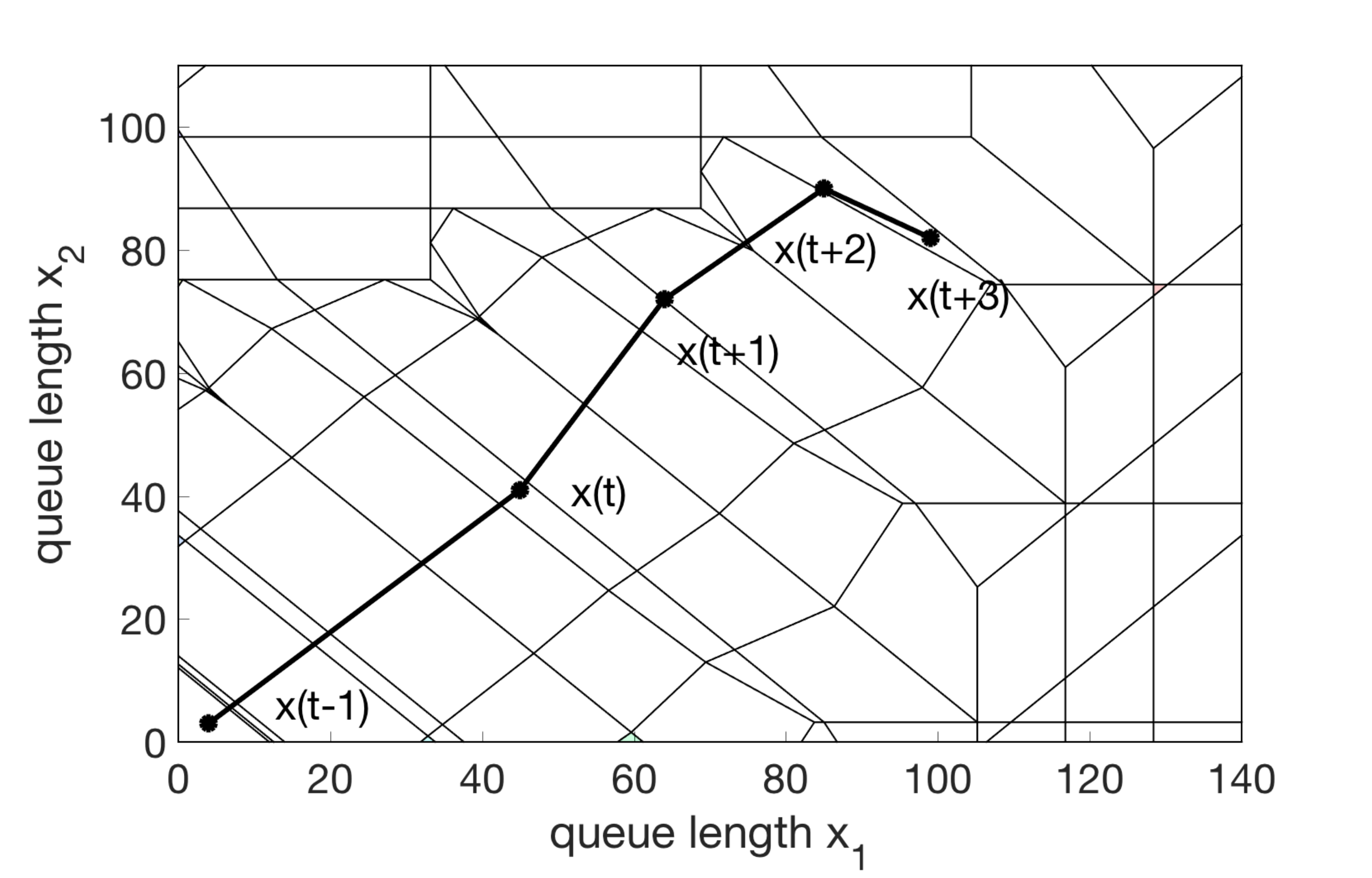}
\caption{Homotopy paths from one QP to the next across multiple state regions: the real line is the homotopy path created by online active set strategy, where $x(t)$ denotes the sampled state within $t^{th}$ control cycle.}\label{fig:OAS}
 \end{figure}

    OASS is an effective method to solve QPs sequently, because it can exploit the geometrical property of the MPC problem.
    For solving the current problem $QP(x(t))$, the OASS moves on a straight line in the parameter space from the previous one $QP(x(t-1))$ as Figure \ref{fig:OAS}.
    This figure shows the parametric space of the two-dimensional traffic example introduced in the section 2, and uses a trajectory to indicate the  dynamic change of the queue length in each road.
    Note that once the trajectory across the boundary of a state region, it means an active set change happened.
    The number of active set changes required by MPC with online active set strategy are reduced to 7, 3, 4, 1, which shows that the online computation complexity have been significantly reduced.

To achieve this idea, the definitions are given as follows.

    \begin{subequations}\label{changshuupdate}
    \begin{gather}
    x_0=x(t-1),\\
    x_0^{new}=x(t),\\
    \bigtriangleup x_0:=x_0^{new}-x_0,\\
    \bigtriangleup g:=g(x_0^{new})-g(x_0)=F^T\bigtriangleup x_0,\\
    \bigtriangleup b:=b(x_0^{new})-b(x_0)=E\bigtriangleup x_0.
    \end{gather}
   \end{subequations}
   gradient and constraint vector are re-parameterised as follows:
   \begin{subequations}
   \begin{gather}
    \tilde{x_0}:[0,1]\to \mathbb{R}^n,\ \tilde{x_0}(\tau):=x_0+\tau\bigtriangleup x_0\\
    \tilde{g}:[0,1]\to \mathbb{R}^n,\ \tilde{g}(\tau):=g(x_0)+\tau\bigtriangleup g\\
    \tilde{b}:[0,1]\to \mathbb{R}^s,\ \tilde{b}(\tau):=b(x_0)+\tau\bigtriangleup b
    \end{gather}
   \end{subequations}
   We assume that the starting point is the known optimal solution $U_0^*$ and $\lambda_0^*$ (and corresponding working set $\mathbb{A})$ of the last $QP(x_0)$ and want to solve $QP(x_0^{new})$.
   The basic idea of the online active set strategy is to move from $x_0$ towards $x_0^{new}$, and thus from $(U^*,\lambda^*)$ towards $(U_{new}^*,\lambda_{new}^*)$, while keeping primal and dual feasibility, $i.e.$, optimality, for all intermediate points. This means that we are looking for homotopies($\mathbb{M}:={1,...,m}$)
   \begin{subequations}\label{direction}
   \begin{gather}
    \tilde{U}^*:[0,1]\rightarrow\mathbb{R}^n,\ \tilde{U}^*(0)=U^*,\tilde{U}^*(1)=U_{new}^*,\\
    \tilde{\lambda}^*:[0,1]\rightarrow\mathbb{R}^n,\ \tilde{\lambda}^*(0)=\lambda^*,\tilde{\lambda}^*(1)=\lambda_{new}^*,\\
    \tilde{\mathbb{A}}:[0,1]\rightarrow 2^{\mathbb{M}},\ \tilde{\mathbb{A}}(0)=\mathbb{A},\tilde{\mathbb{A}}(\tau)\subseteq \mathbb{M},\\
    \tilde{I}:[0,1]\rightarrow 2^{\mathbb{M}},\ \tilde{I}(\tau):=\mathbb{M}\setminus\tilde{\mathbb{A}}(\tau).
   \end{gather}
   \end{subequations}
   which satisfy the well-known KKT conditions at every point $\tau\in[0,1]$:
   \begin{subequations}\label{KKT}
   \begin{gather}
   \left(\begin{array}{cc} H & G_{\tilde{A}(\tau)}^T\\G_{\tilde{A}(\tau)} & 0\end{array} \right)\left(\begin{array}{c}\tilde{x}^*(\tau)\\-\tilde{\lambda}^*_{\tilde{\mathbb{A}}(\tau)}\end{array}\right) = \left(\begin{array}{c}-\tilde{g}(\tau)\\\tilde{b}_{\tilde{\mathbb{A}}(\tau)}(\tau)\end{array}\right), \\
   \tilde{\lambda}_{\tilde{\mathbb{I}}(\tau)}^*(\tau)=0,\\
   G_{\tilde{\mathbb{I}}(\tau)}\tilde{x}^*(\tau)\geq b_{\tilde{\mathbb{I}}(\tau)}(\tau),\\
   \tilde{\lambda}_{\tilde{\mathbb{A}}(\tau)}^*\geq 0.
   \end{gather}
    \end{subequations}

    Since $\tilde{U}^*(\tau)$ and $\tilde{\lambda}^*(\tau)$ are piecewise linear functions, as already shown in \cite{Bemporad2002}, locally a relation of the form
    \begin{subequations}\label{update}
    \begin{gather}
    \tilde{U}^*(\tau) = U^* + \tau \bigtriangleup x^*,\\
    \tilde{\lambda}^*_{\mathbb{A}}(\tau) = \lambda_{\mathbb{A}}^* + \tau \bigtriangleup \lambda_{\mathbb{A}}^*.
    \end{gather}
    \end{subequations}
    holds for sufficiently small $\tau\in [0,\tau_{max}]$, $\tau_{max}\geq 0$.

    Conditions (\ref{KKT}) are met at $\tau=0$, as we start from an optimal solution. Therefore equality (\ref{KKT}a) is satisfied for all $\tau \in [0,\tau_{max}]$ if and only if
    \begin{equation}\label{18a}
    \left(\begin{array}{cc}H & G_{\mathbb{A}}^T\\G_{\mathbb{A}} & 0\end{array}\right)
    \left(\begin{array}{c}\bigtriangleup U^* \\ -\bigtriangleup \lambda_{\mathbb{A}^*}\end{array}\right)\\
    =\left(\begin{array}{c}-\bigtriangleup g\\ \bigtriangleup b_{\mathbb{A}}\end{array}\right)
    \end{equation}
    holds.
    Equation (\ref{18a}) has a unique solution, as long as $G_{\mathbb{A}}$ is row full rank which can  be  easily ensured.

    As long as stays within a critical region,\footnote{The parameter space $\mathbb{P}$ can be divided into (convex sets, so-called critical regions. Such that the QP solution for all $x_0$ within one critical region has an identical optimal active set.)} the QP solution depends on $x_0$, but it might happen that one has to cross the boundaries of critical regions during the way along the straight line.
    The active set stays constant as long as no previously inactive constraint become active, $i.e.$,
    \begin{equation}
    G_i^T(x^*+\tau\bigtriangleup x^*) = b_i(w_0) + \tau\bigtriangleup b_i \nonumber
    \end{equation}
    for some $i\in\mathbb{I}$, and no previously active constraint becomes inactive, $i.e.$,
    \begin{equation}
    \lambda_i^* + \tau \bigtriangleup \lambda_i = 0
    \end{equation}
    for some $i\in \mathbb{A}$.
    Thus, the maximum possible homotopy step length $\tau_{max}$ is determined as follows:
    \begin{subequations}\label{tau_max}
    \begin{gather}
    \tau_{max}^{prim}:=\min_{{i\in\mathbb{I},{G^T_i\bigtriangleup U^* < \bigtriangleup b_i}}}\frac{b_i(x_0)-G_i^TU^*}{G^T_i\bigtriangleup U^* - \bigtriangleup b_i}\\
    \tau_{max}^{dual}:=\min_{i\in \mathbb{A},\bigtriangleup \lambda_i < 0}-\frac{(\lambda^*)_i}{\bigtriangleup \lambda_i},\\
    \tau_{max}=\{\min{1,\tau_{max}^{prim},\tau_{max}^{dual}}\}
    \end{gather}
    \end{subequations}

    In the case that $\tau_{max}$ equals one, the new state $x_0^{new}$ has been reached, and the solution of the new quadratic program $QP(x_0^{new})$ is found at meantime.
   In the other cases, a constraint will be removed or added to the active set $\mathbb{A}$ which limit $\tau_{max}$ to reaching 1.
   After updating the active set, the whole procedure repeats again, and a new step direction and length are obtained.
   This iteration stop until the $\tau_{max}$ equal to one, $i.e.$, the solution of $QP(x_0^{new})$ is found.
   Online active set strategy is summarized in Algorithm \ref{alg:oass}
    \begin{algorithm}[htb]
      \caption{ Online active set strategy (OASS).}
      \label{alg:oass}
      \begin{algorithmic}[1]
        \REQUIRE
          Solution $(U^*_{i-1},\lambda^*_{i-1})$ of $QP(x_{i-1})$, corresponding working set $\mathbb{A}$, new parameter $x_{i}^{new}$
        \ENSURE
          Solution $(U_{i}^*,\lambda_{i}^*)$ of $QP(x_0^{i})$, corresponding working set $\mathbb{A}^{i}$
        \STATE Calculate $\bigtriangleup x_0$, $\bigtriangleup g$ and $\bigtriangleup b$ via equation (\ref{changshuupdate}).
        \STATE Calculate primal and dual step directions $\tilde{U^*}$ and $\tilde{\lambda^*}$ via equation (\ref{direction}).
        \STATE Determine maximum homotopy step length $\tau_{max}$ from equation (\ref{tau_max}).
        \STATE Obtain optimal solution of $QP(\tilde{x}_0)$:\\
        \qquad $\tilde{U}^*(\tau) = U^* + \tau \bigtriangleup x^*$\\
        \qquad $\tilde{\lambda}^*_{\mathbb{A}}(\tau) = \lambda_{\mathbb{A}}^* + \tau \bigtriangleup \lambda_{\mathbb{A}}^*$\\
        \STATE if $\tau_{max} = 1$:\\
            \qquad Set $U^*_{i}\leftarrow \tilde{U}^*, \lambda^*_{i}\leftarrow \tilde{\lambda}^*, \mathbb{A}^{i}\leftarrow\mathbb{A}$. stop!
        \STATE if $\tau_{max} = \tau_{max}^{dual}$:\\
            \qquad $\mathbb{A}\leftarrow \mathbb{A}\setminus\{j\} (\tau_{max}^{dual}=-\frac{(\lambda^*)_j}{\bigtriangleup \lambda_j})$
        \STATE if $\tau_{max} = \tau_{max}^{prim}$:\\
            \qquad $\mathbb{A}\leftarrow \mathbb{A}\cup\{j\} (\tau_{max}^{prim}=\frac{b_j(x_0)-G_j^TU_0^*}{G^T_j\bigtriangleup U_0^* - \bigtriangleup b_j})$
        \STATE Set $x_0\leftarrow \tilde{x_0}, U^*\leftarrow \tilde{U}^*, \lambda^*\leftarrow \tilde{\lambda}^*$, continue with step (2)
      \end{algorithmic}
    \end{algorithm}

\subsection{Real-time variant for traffic signal control problem}
  
    One advantage of online active set strategy is that it produces a sequence of optimal solutions for QPs on the homotopy path. 
Thus, it is possible to interrupt this sequence after every partial step and start a new homotopy from the current iterate towards the next QP\cite{Ferreau2008}.

This advantage inspires us to propose a new framework of  applying MPC to traffic signal control problem.
The basic idea of our framework is  to divide the cycle time $T_c$ into several sample intervals $T_s$.
Online active set strategy outputs the optimal solution every other sampling interval until the last interval of the control cycle.
Since online active set strategy can veer to new homotopy path during the iteration process, the solution is gradually approaching suboptimal solution of the current cycle.   
Based on this framework, the computation load in the end of each cycle is distributed to several sampling intervals.

The most important part of this framework is to choose the sample interval length $T_s$.
One naive approach to determine $T_s$ would be directly divide $T_c$ into arbitrary number of intervals.
However, this approach ignores the fact that if $T_s$ is too large, the computation complexity, $i.e.$ the number of active set change, in the end of the cycle can still be unbearable.
Instead, if $T_s$ is too small, it will cause great sample complexity.
For the purpose of allocating the computation complexity to every sample interval,
we choose $T_s$ in terms of the positive correlation between the number of active set changes and CPU time.
To be more specific, if one obtains an estimate for the number of active set changes $n_a$ from last cycle to the next, $e.g.$ using closed-loop simulations, it is easy to estimate the possible sampling time length and the number of sample interval $n_{itr}$.
\begin{equation}
T_s = \frac{T_c}{n_a+1}
\end{equation}
\begin{equation}
n_{itr} = n_a+1
\end{equation}

Even thought, we choose $T_s$ based on the average number of active set changes which could be extremely inaccurate in some cases, because of the stochastic of traffic system.
This framework is still robust to external disturbances.
Since if too many active set changes are necessary to get from the solution of the old QP to that of the current QP we just stop the solution of the current QP and start a new homotopy towards the solution of the next QP.

Our new traffic signal control framework is summarized in Algorithm \ref{alg:framework}.

 \begin{algorithm}[htb]
      \caption{}
      \label{alg:framework}
      \begin{algorithmic}[1]
        \REQUIRE
          Solution $(U^*_0,\lambda^*_0)$ of $QP(x_0)$, corresponding working set $\mathbb{A}$, new parameter $x_0^{new}$, sample interval number $n_{iter}$
        \ENSURE
          Solution $(U_{new}^*,\lambda_{new}^*)$ of $QP(x_0^{new})$, corresponding working set $\mathbb{A}^{new}$
        \FOR{$i=1$; $i\le n_{iter}$; $i++$ }  
    \STATE Obtain the new parameter $x_0^{i}$ 
     \STATE $(U_i^*,\lambda_i^*)=OASS(U_{i-1}^*,\lambda_{i-1}^*,x_0^i, n_{iter})$:\\
  \ENDFOR
  \STATE $(U_{new}^*,\lambda_{new}^*) = (U_{n_{iter}}^*,\lambda_{n_{iter}}^*)$
      \end{algorithmic}
    \end{algorithm}
    
  \begin{figure}[htbp]
        \centering
        \includegraphics[width=0.80\textwidth]{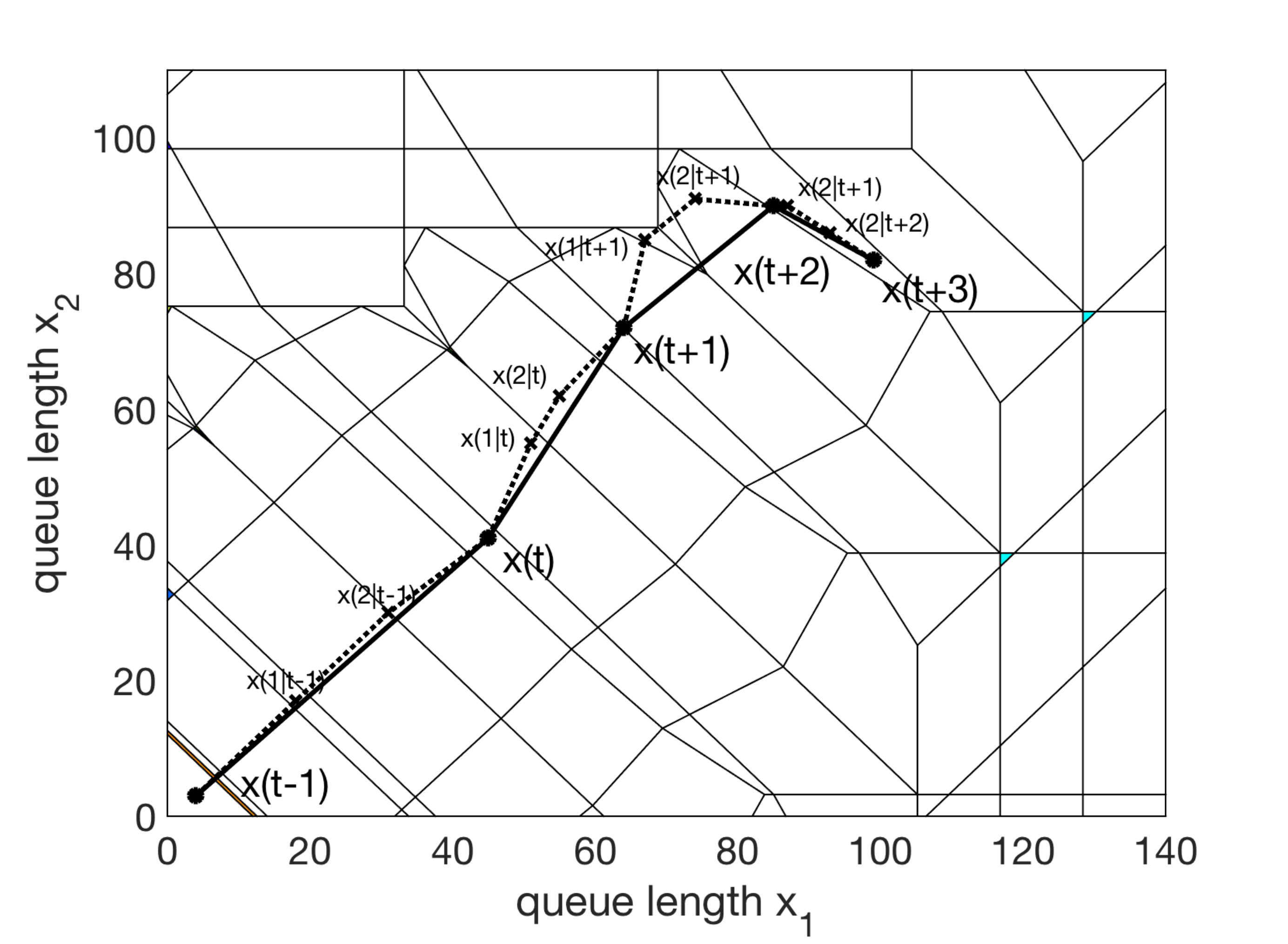}
        \caption{Homotopy paths from one QP to the next across multiple state regions: the doted line is the homotopy path created by online active set strategy under the new real-time framework, where $x(i|t)$ denotes the $i^{th}$ sampled state within $t^{th}$ control cycle.}\label{fig:our_parm}
    \end{figure}
    

To be more comprehensible, the 2D traffic system example introduced above  will serve to this idea.
The corresponding polyhedral partition of the state-space which has 44 polyhedral cells and the state dynamic trajectory are depicted in Figure  \ref{fig:our_parm}. 
The number of active set changes in each control cycle has been distributed to each sample intervals including middle sample intervals (not last ones of each control cycle), and  last sample intervals.
Since, in the last sample interval, the system has to perform the optimal signal strategy, the real-time feasibility in this  interval is the mainly focus. 
The number of active set changes in each last sample interval are 1, 1, 1, 0, which shows a great improvement in real-time performance of traffic control system.

\begin{remark}
Although, under the proposed framework, the total number of active set change slightly increased in some control cycle,  changes is reduced to a an extremely small amount at the last interval of each cycle, which is more practical for the traffic-response based traffic control system.
\end{remark}    
    
\begin{remark}
In each sample interval, the solution $(U^{*}_{i},\lambda^{*}_{i})$(except $i=iter$) won't be applied to traffic system.
Then, the MPC controller  have plenty of time to finish the online active strategy.
In most cases, there will no any active set change in the last sample interval, which means our framework can give the same green time plan  as MPC with a extremely small delay. 
\end{remark}

\section{Simulation}
\subsection{Roads networks modeling}  
\begin{figure}[htbp]
        \centering
        \includegraphics[width=\textwidth]{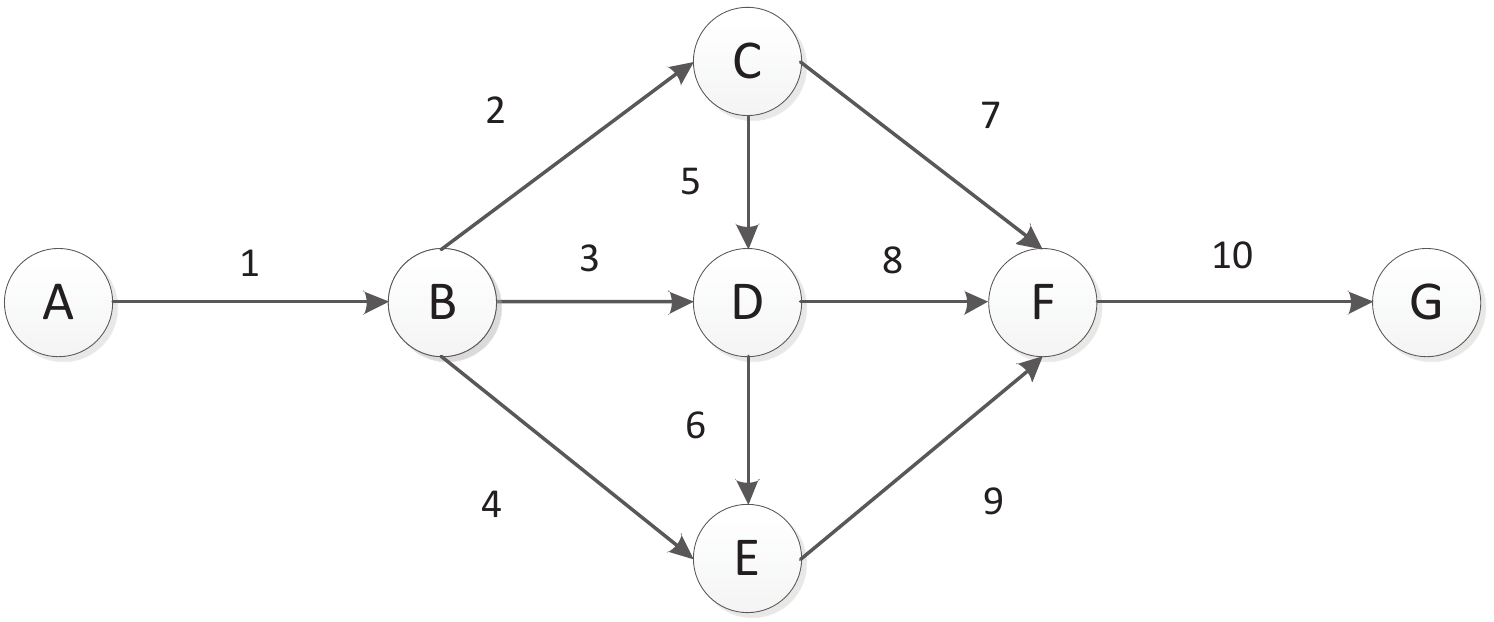}
        \caption{A toy traffic network for simulation experiments.}
        \label{fig: ToyNetwork}
    \end{figure}

To preliminarily investigate the comparative efficiency and real-time feasibility of the developed approaches to the problem of traffic signal control, the toy traffic network is considered. 
Our toy traffic network model, shown as in Figure \ref{fig: ToyNetwork} was created by the traffic network modeler, Paramics.

    As the Figure \ref{fig: ToyNetwork} shows, the traffic network consists of 5 signalized junctions $B,C,D,E,F$, 10 one-way links whose capacity are 100.
    The node $A$ and $G$ is defined as the source node and sink node respectively.
    Other basic parameters set for our simulation experiments are listed in Table \ref{tab:trafficpara}.
   
\begin{table}[htbp] 
\centering
 \caption{Traffic system parameters} 
 \label{tab:trafficpara}
\begin{tabular}{cccc}
  \toprule 
  Parameters & Physical meaning & Values\\
  \midrule
$T_c$&Cycle time& 55s\\
$T_{horizon}$ & Simulative control horizon& 6000s(100 cycles)\\
$N$ & Prediction horizon & 180s (3 cycles)\\
$g_{min}$ & Minimum of green time & 5s\\
$g_{max}$ & Maximum of green time & 55s\\
$t_s$ & Simulation time step & 0.1s\\
$q_s$&Saturation flow rate& 1900veh/h\\
$\tau_{left}$& The probability a vehicle turns left & 30\%\\
$\tau_{right}$& The probability a vehicle turns left & 30\%\\
$\tau_{straight}$& The probability a vehicle goes straight & 40\%\\
\bottomrule 
 \end{tabular} 
\end{table}
   
\subsection{Simulation approach}
 \begin{figure}[htbp]
        \centering
       \includegraphics[width=\textwidth]{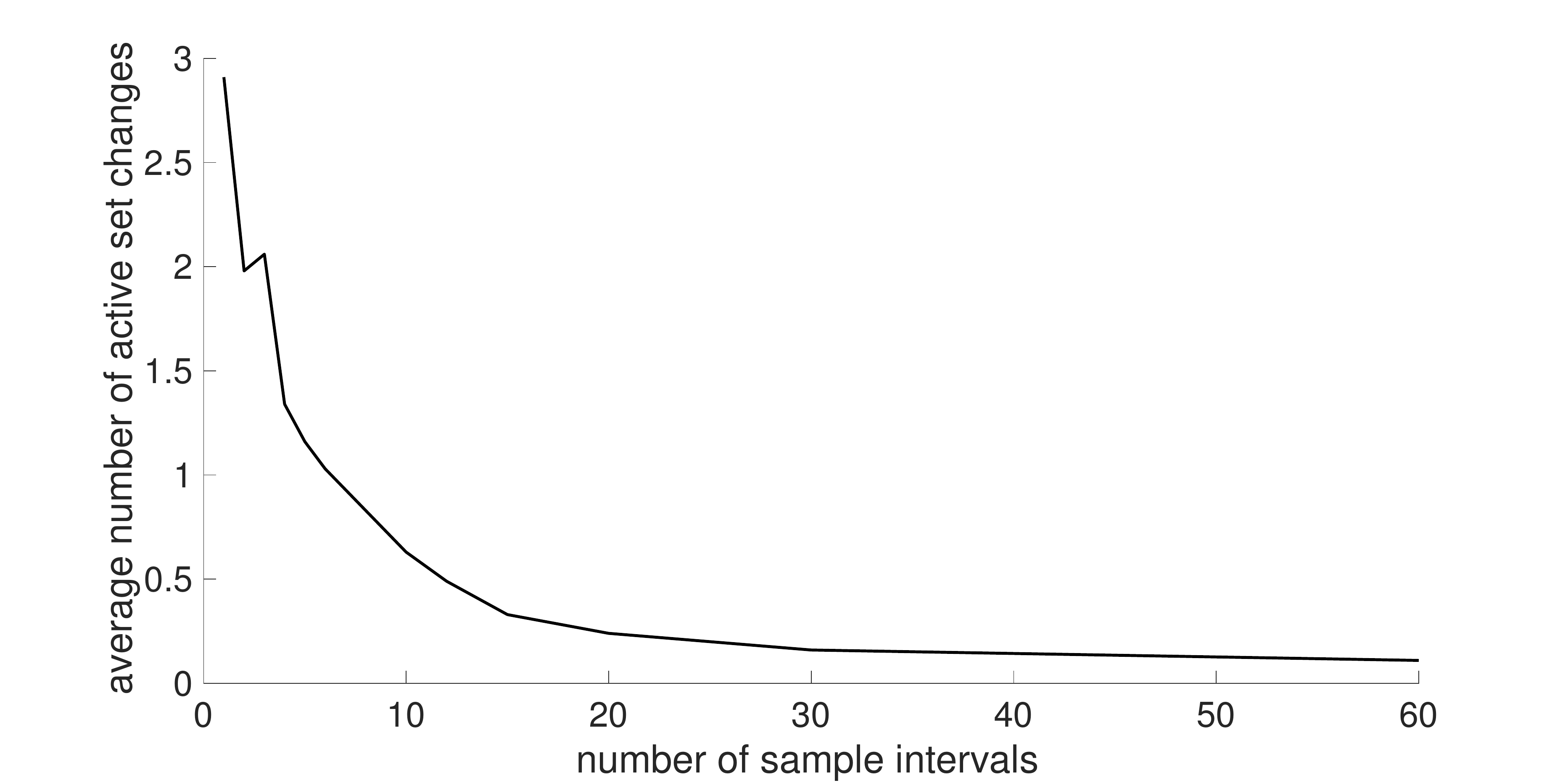}
        \caption{The average number of active set changes required by MPC-ours along the increment of sample interval number (in the constant scenario).}
             \label{fig:samplelen}
    \end{figure}
We used traffic network modeler Paramics to simulate the unban traffic system based on the parameters as shown in Table \ref{tab:trafficpara}. 
Due to long developing time and heavy computational efforts, the OASS algorithm is achieved in Matlab using a qbOASS toolbox \cite{Ferreau2014}.
When the Paramics is working, the programs which capture the traffic information, output the optimal solution, and implement the optimal signal split strategy will be loaded into Paramics as a plug-in module in DLL form.

Several tests were conducted in order to investigate the behavior of the three alternative methodologies.
    
    \textbf{MPC}: MPC with traditional active set strategy
    
    \textbf{MPC-oass}: MPC with online active set strategy as presented in Section 3.1, $i.e.$, the solver is initialized with the solution of the previous QP of last control cycle.
    
    \textbf{MPC-ours}: MPC with the new calculation framework based on the online active set strategy as presented in Section 3.2, $i.e.$, the solver is initialized with the solution of previous sample interval, and only the solution of last sample interval will be implemented into the traffic system.  

Two traffic scenarios are considered:

\textbf{Constant scenario}: The inflow rate of the source node where vehicle drive into the controlled traffic network is stay constant (1200 veh/h) during  the simulation.

\textbf{Random scenario}: The inflow rate of the source node varies between 200 veh/h and 2400 veh/h which stayed constant during one cycle time, and was randomly selected in the next.

We have used the number of active set changes and CPU time spend to assess the effectiveness of our proposed control framework in each scenario.
Since, for traffic control system, the real-time constraint is on the each end of control cycle.
It is reasonable that, for MPC-ours, we only compare the change numbers in the last sample interval. 

\subsection{Simulation results}
      \begin{figure}[htbp]
        \centering
       \includegraphics[width=\textwidth]{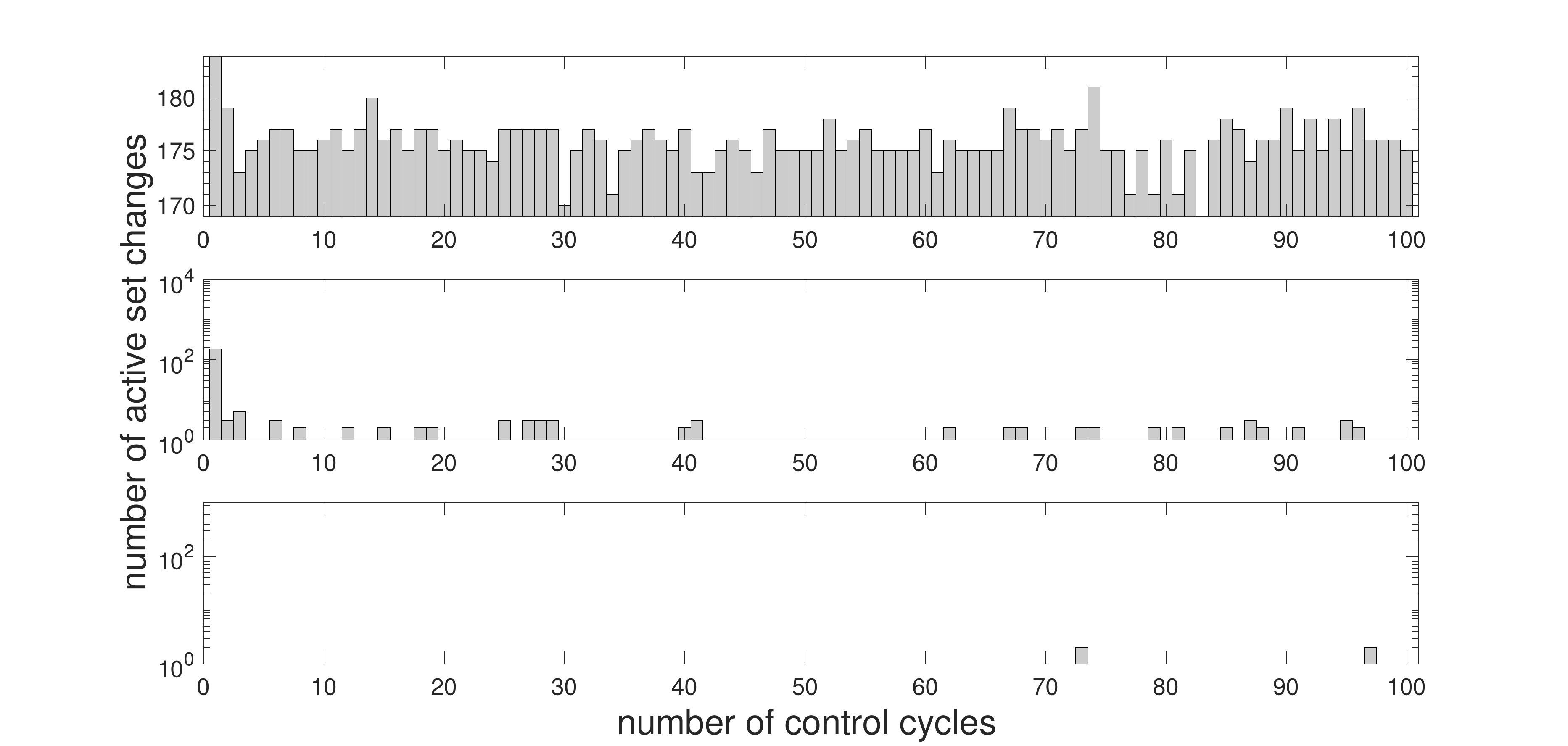}
        \caption{Constant scenario: number of active set changes required by standard QP solver with MPC, MPC-oass, MPC-ours respectively .}
             \label{fig:NC}
    \end{figure}
    In order to demonstrate the computational savings as a result of the discussion in Section 3, firstly the comparison is made on the number of active set changes necessary for the algorithm to converge to the optimal solution at every control cycle.
    
    Figure \ref{fig:samplelen} illustrates the average number of active set changes when the scaling parameter $n_s$(number of sample interval) varies from 1 to 60. 
    Roughly speaking, one can imagine that a good choice of scaling factor in this case is around 30, which makes an appropriate trade-off between online processing and sample complexity.
    Therefore, in the following experiment, $n_s$ is set to 30 in each control cycle. 

\begin{figure}[htbp]
        \centering
       \includegraphics[width=\textwidth]{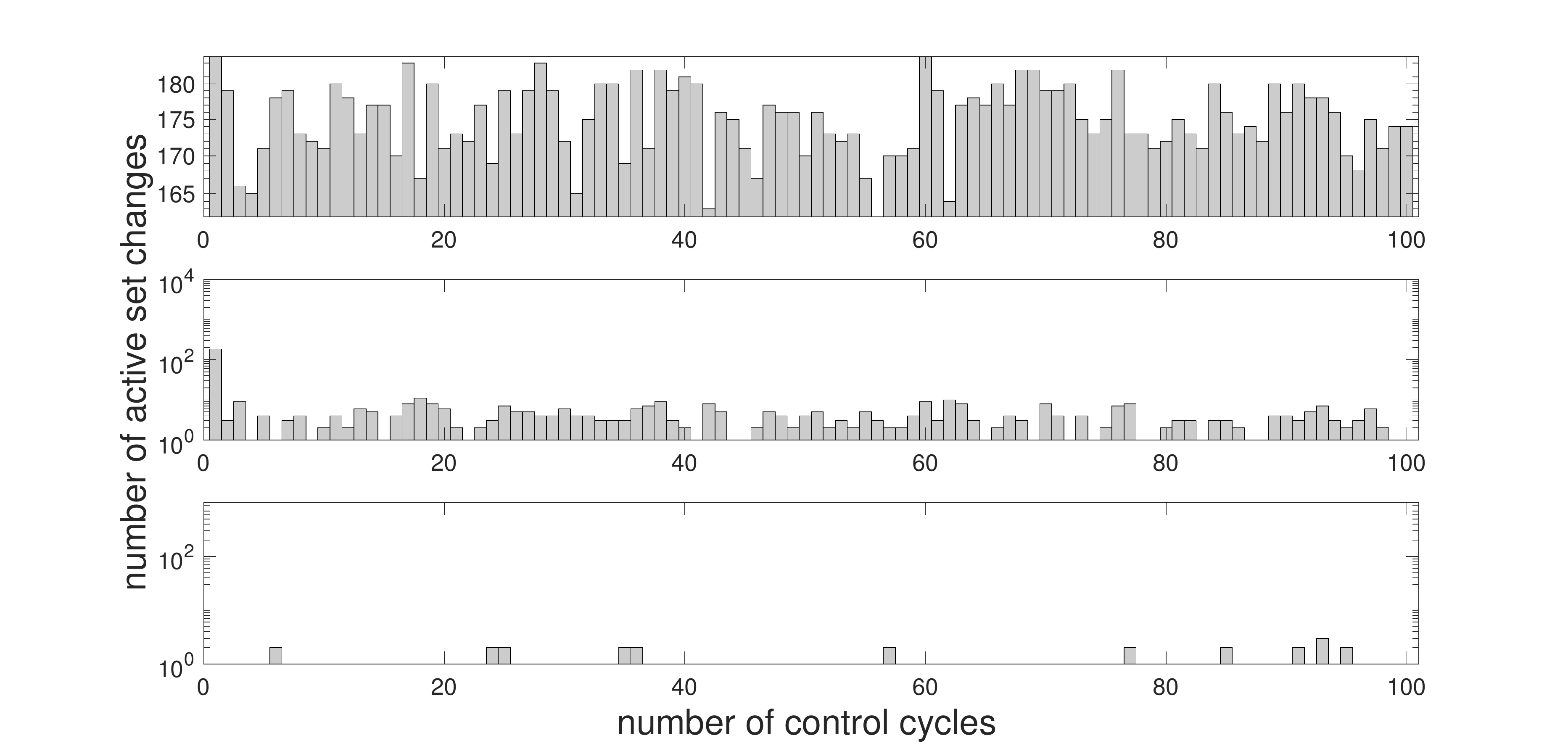}
        \caption{Random scenario: number of active set changes required by standard QP solver with MPC, MPC-oass, MPC-ours  respectively .}
        \label{fig:UNC}
    \end{figure}

    Figure \ref{fig:NC},\ref{fig:UNC} depicts the number of active set changes for methods MPC, MPC-oass and MPC-ours in the both scenarios.

    In the constant scenario, as can be seen in Figure \ref{fig:NC}, with a great initial start point MPC-ours and MPC-oass does not need any active set changes in most of the cycle.
    It was due to the solution of last sample interval lies in the same parametric region as the optimal solution.
    However, since MPC  didn't give estimate of the optimal solution, it performed more active set changes towards the new solution.
    
    In the random scenario, sometimes, several links have much space while are relatively crowned in the other cycles.
    This situation lead to vehicles have more freedom which means the state of traffic could change dramatically.
    Therefore, as can be seen in Figure \ref{fig:UNC}, the performance of MPC-oass is worse than previous scenario. 
    In contrast, MPC-ours is still able to identify the active constraints at optimality for most of cycles.
    As for MPC, because the initial solution is random, its number of active set changes is uncontrollable in both of the scenarios.

Next, we have listed the CPU time spend of each method in Table \ref{tab:test}.
These results show that both the MPC-oass and MPC-ours controller can greatly improve the real-time performance of MPC-based traffic signal control system.
\begin{table}[htbp]
    \centering
 \caption{\label{tab:test}Maximum and average CPU time spend [ms] when solving the QP problems with different strategies}
 \begin{tabular}{ccccc}
  \toprule
  Strategies &   \multicolumn{2}{c}{Constant}    &  \multicolumn{2}{c}{Random} \\
  \cline{2-3}  \cline{4-5}
    ~                                                   & MaxTime & AvgTime & MaxTime & AvgTime\\
  \midrule
     MPC    &  1024      & 835       & 1010    & 861 \\
    MPC-oass   &  14       & 0.4370  & 22      &0.49       \\
    MPC-ours    &  0.22      & 0.1035       & 0.4380     & 0.1537 \\
    \bottomrule
 \end{tabular}
\end{table}
 
In order to further validate the effectiveness of MPC-ours, we illustrated the computation time improvement of the MPC-ours controller with respect to MPC-oass controller in Table \ref{tab:our_oass}. 
\begin{table}[htbp]
    \centering
 \caption{CPU time spend comparison of MPC-ours with respect to MPC-oass ($\rho$ = AvgTime of MPC-ours/AvgTime of MPC-oass)}
 \label{tab:our_oass}
 \begin{tabular}{cccc}
  \toprule
Time comp.                                     & \multicolumn{2}{c}{Scenarios} \\
 ~                                              & Constant  & Random \\
  \midrule
$\rho \leq .5$                 & 27\%      & 56\%       \\
$.5 < \rho \leq 1$  & 59\%     & 25\%      \\
$1 < \rho \leq 1.5$ & 12\%     & 5\%      \\
$1.5 < \rho$                 & 2\%      & 4\%      \\
  \bottomrule
 \end{tabular}
\end{table}

    We defined the comparative index $\rho$ to be the ratio of the CPU time using of strategy MPC-ours to the CPU time using of strategy MPC-oass.
    For each scenario type, we computed the percentage of 100 QP problems (the control horizon is 100 control cycles) for which $\rho \leq .5$(MPC-ours is 'much' better than MPC-oass), $.5 \leq \rho \leq 1$ (MPC-ours is 'better' than MPC-oass), $1 \leq \rho \leq 1.5$ (MPC-ours is 'worse' than MPC-oass), and $1.5 \leq \rho$ (MPC-ours is 'much worse' than MPC-oass).
       
    A careful examination of Table \ref{tab:our_oass} reveals that MPC-ours usually performed at least as well as MPC-oass in the two scenarios.
    More specifically, the percentages reported in the last two rows both are really small.
    Table \ref{tab:test} also illustrates that the performance of MPC-oass usually degraded in the random scenario, which is consistent with the conclusion obtained by the comparison of active set changes.
    As can be seen, the percentage of 'much better' increase greatly in the random scenario.
    This is an expected result, as larger randomness of inflow rate tend to increase the distance between the original solution of previous cycle  and the new one, which  violated the applicable condition of OASS algorithm \cite{Ferreau2008}.
    Therefore, the advantages of MPC-oass are less pronounced.
    Table \ref{tab:our_oass} also indicates that MPC-ours usually results in the larger savings in terms of CPU time.
    Specifically, MPC-our almost always has the larger percentage in the 'much better' and 'better' rows.

\section{Conclusion}
    In this paper, we presented an efficient framework for solving the sequential quadratic programming problem from the analysis of the application of model predictive control for controlling traffic network.
    Firstly, we proposed the load unbalance problem in traffic control system which is based on traffic-responsive algorithm.
    For avoiding this problem and improve the real-time performance of MPC-based traffic signal control system, we introduced online active set strategy which take full advantage of the parametric nature of MPC problems to accelerate the solution of QP problem  .
    Furthermore, we demonstrated a  way to distribute the computation complexity into entire time step by divide the control cycle into several sample intervals, and apply online active set strategy at every interval.
    It was shown that our new framework can achieve a significantly reduction in computational cost, when compared to other methods, which make it possible to be used for controlling larger urban traffic network.
    Further research will focus on developing this framework  to deal with more complex control objections in real-world conditions, as well as utilize more elaborated nonlinear flow model to improve the control performance.

\bibliographystyle{unsrt}
\bibliography{bibtex}

\end{document}